\documentclass[twocolumn]{aastex631}
\usepackage{import}
\usepackage{graphicx}
\usepackage[caption=false]{subfig}
\usepackage{verbatim}
\usepackage{CMNDSManual}
\usepackage{CMNDSCont}
\usepackage{CMNDSCorrCont}
\usepackage{CMNDSSampSZ}
\usepackage{CMNDSMBHScalRel}
\usepackage{CMNDSMasses}
\usepackage{CMNDSLXMstar}
\usepackage{CMNDSFEddScalRel}
\usepackage{CMNDSMstarDeltaS}
\usepackage{CMNDSDeltaSPetroRad}
\usepackage{CMNDSBinary}

\shorttitle{HLX Stellar Cores}
\shortauthors{Barrows et al.}

\begin{document}

\accepted{for publication in ApJ}

\title{Merger-Driven Growth of Intermediate-mass Black Holes: Constraints from \emph{Hubble Space Telescope} Imaging of Hyper-luminous X-Ray Sources}

\author[0000-0002-6212-7328]{R. Scott Barrows}
\affiliation{Department of Astrophysical and Planetary Sciences, University of Colorado Boulder, Boulder, CO 80309, USA}

\author[0000-0003-4440-259X]{Mar Mezcua}
\affiliation{Institute of Space Sciences (ICE, CSIC), Campus UAB, Carrer de Magrans, E-08193 Barcelona, Spain}
\affiliation{Institut d’Estudis Espacials de Catalunya (IEEC), Carrer Gran Capità, E-08034 Barcelona, Spain}

\author{Julia M. Comerford}
\affiliation{Department of Astrophysical and Planetary Sciences, University of Colorado Boulder, Boulder, CO 80309, USA}

\author[0000-0003-2686-9241]{Daniel Stern}
\affiliation{Jet Propulsion Laboratory, California Institute of Technology, 4800 Oak Grove Drive, Pasadena, CA 91109, USA}

\correspondingauthor{R. Scott Barrows}
\email{Robert.Barrows@Colorado.edu}

\begin{abstract}

Hyper-luminous X-ray sources (HLXs) are extragalactic off-nuclear X-ray sources with luminosities exceeding the theoretical limit for accretion onto stellar-mass compact objects. Many HLXs may represent intermediate-mass black holes (IMBHs) deposited in galaxy halos through mergers, and properties of the stellar cores surrounding HLXs provide powerful constraints on this scenario. Therefore, we have systematically built the largest sample of HLX candidates with archival \emph{Hubble Space Telescope} (\hst) imaging (\Samp) for the first uniform population study of HLX stellar cores down to low masses. Based on their host galaxy redshifts, at least 21 (88\%) have stellar core masses $\geq$\,$10^7$\,\Msun~and hence are consistent with accretion onto massive black holes from external galaxies. In 50\% of the sample, the \hst~imaging reveals features connecting the HLXs with their host galaxies, strongly suggesting against the background/foreground contaminant possibility in these cases. Assuming a mass scaling relation for active galactic nuclei and accounting for an estimated contamination fraction of \PercCorrCont\%, up to $\sim$60\% of our sample may be associated with IMBHs. Similar to previously known HLXs, the X-ray luminosities are systematically elevated relative to their stellar core masses, possibly from merger-driven accretion rate enhancements. The least massive stellar cores are preferentially found at larger nuclear offsets and are more likely to remain wandering in their host galaxy halos. The HLX galaxy occupation fraction is $\sim$\,$10^{-2}$ and has a strong inverse mass dependence. Up to three of the HLX candidates (12\%) are potentially consistent with formation within globular clusters or with exceptionally luminous X-ray binaries.

\end{abstract}


\section{Introduction}
\label{sec:intro}

Hyper-luminous X-ray sources (HLXs) are intrinsically luminous X-ray sources ($L_{X}$\,$\ge$\,$10^{41}$\,\uLum) located in off-nuclear regions of galaxies \citep[e.g.,][]{Gao:2003}. In contrast to the population of ultra-luminous X-ray sources (ULXs; $L_{X}$\,$\ge$\,$10^{39-41}$\,\uLum; \citealp[see][and references therein]{Kaaret:2017}) that are typically associated with X-ray binaries (XRBs; e.g., \citealp{Remillard:McClintock:2006}), the higher luminosities of HLXs suggest they are most likely powered by accretion onto compact objects more massive than stars \citep[e.g.,][]{Miller:2004,Miller:Colbert:2004,King:2005}. This distinction is further suggested by a break in the X-ray point source luminosity function at $L_{X}$\,$\sim$\,1\,$-$\,2\,$\times$\,$10^{40}$\,\uLum~\citep[e.g.,][]{Swartz:2011,Mineo:2012,Tranin:2023}.

Observationally, HLXs are remarkable for their faint or undetected optical counterparts when compared to their X-ray luminosities \citep[e.g.,][]{Matsumoto:2001,Farrell:2009,Jonker:2010,Mezcua:2013a,Mezcua:2013c,Kim:2015,Mezcua:2015,Comerford:2015,Lin:2016,Mezcua:2018b,Lin:2020}. The corresponding stellar masses are typically less massive than those of the stellar bulges associated with supermassive black holes (SMBHs; \MBH\,$>10^{6}$\,\Msun) found in the nuclei of massive galaxies. Therefore, many HLXs may be powered by accretion onto intermediate-mass black holes (IMBHs; \MBH\,$=10^{2-6}$\,\Msun; see \citealt{Mezcua:2017} for a review) previously hosted by low-mass galaxies. Indeed, frequent mergers of satellites with massive galaxy halos is expected based on observations of minor merger rates \citep[e.g.,][]{Conselice:2003,Lotz:2011} and of tidal streams in the Milky Way \citep[e.g.,][]{Belokurov:2006} and M31 \citep[e.g.,][]{Ibata:2001,Ferguson:2002}.

While major galaxy mergers (as opposed to minor mergers) are theoretically more efficient at funneling gas and dust toward galaxy nuclei \citep[e.g.,][]{Hernquist:1989,Hopkins2008}, the more frequent occurrences of minor mergers suggest they may have a significant impact on evolution of the black hole mass distribution through merger-driven growth \citep[e.g.,][]{Cox:2008,Fakhouri:2010,Barrows:2017}. Moreover, this process must be accounted for when using IMBHs to constrain formation scenarios of primordial black hole seeds \citep[e.g.,][]{Mezcua:2019}.  In particular, if a large fraction of the current IMBH mass distribution was accreted through mergers, then the true primordial seed masses would have been lower, lending more favor to formation scenarios within dense stellar clusters \citep[e.g.,][]{Zwart:2002} or from Population III stellar remnants \citep[e.g.,][]{Volonteri:2010}, rather than from collapse of pristine primordial gas clouds \citep[e.g.,][]{Lodato:2006}.

Though the accreting sources of some HLXs are strongly suspected to be IMBHs \citep[e.g.,][]{Farrell:2009,Lin:2016,Lin:2020}, the true nature of many other HLXs is unclear. For example, significant tidal stripping of massive galaxy cores can potentially result in the low-mass stellar cores hosting some HLXs, resembling compact dwarf galaxies \citep[e.g.,][]{Seth:2014}. Alternatively, HLXs could possibly be associated with IMBHs that formed in-situ within the host galaxy in dense stellar clusters \citep[e.g.,][]{Zwart:2002,Mapelli:2016}. As another alternative,  in rare cases super-Eddington accretion onto compact stellar-mass objects has been observed to produce X-ray luminosities near $10^{41}$\,\uLum~\citep[e.g.,][]{Israel:2017}.

Understanding the nature of HLXs and placing them in the context of the massive black hole population requires estimates of their stellar core masses to determine if they are consistent with galaxy stellar cores, if they show morphological evidence of mergers, and if the mergers elevate their accretion rates. However, only a handful of HLXs have been studied with this level of detail \citep[e.g.,][]{Matsumoto:2001,Farrell:2009,Jonker:2010,Mezcua:2013a,Mezcua:2013c,Kim:2015,Mezcua:2015,Comerford:2015,Lin:2016,Mezcua:2018b,Lin:2020}, and they form a heterogeneous sample. Several previous studies have examined HLXs as a population \citep{Zolotukhin:2016,Gong:2016,Barrows:2019,Tranin:2023}, but the use of ground-based imaging introduces biases toward relatively high-mass stellar cores and/or against the detection of faint morphological features indicative of mergers.

In \citet{Barrows:2019} we previously developed a large catalog of HLX candidates based on X-ray sources from the \ch~Source Catalog~\citep[CSC;][]{Evans:2010} in galaxies from the Sloan Digital Sky Survey (SDSS). To overcome the limitations described above, in this work we assemble archival \emph{Hubble Space Telescope} (\hst) imaging for a subset of this sample with the intention of using the high resolution and high sensitivity data to isolate stellar cores of the HLXs, measure their masses, and search for evidence of interaction with their host galaxies. The efficient sample selection, deep imaging sensitivity, and high spatial resolution will enable the first systematic and uniform study of HLX candidate stellar core morphologies down to low masses, thereby yielding strong constraints on the primary nature of HLXs as a population.

This paper is structured as follows: in Section \ref{sec:proc} we describe our procedure for building the sample, in Section \ref{sec:nature} we discuss implications for the nature of the accreting X-ray sources, in Section \ref{sec:connection} we discuss our sample in the context of merger-driven massive black hole growth, in Section \ref{sec:Occ} we estimate the HLX occupation fraction in galaxies and its dependence on stellar core mass, and in Section \ref{sec:conc} we present our conclusions. Throughout we assume a flat cosmology defined by the nine-year Wilkinson Microwave Anisotropy Probe observations \citep{Hinshaw:2013}: $H_{0}$\,$=$\,69.3\,km\,Mpc$^{-1}$\,s$^{-1}$ and $\Omega_{M}$\,$=$\,0.287.

\begin{figure*}[ht!]
\includegraphics[width=0.96\textwidth]{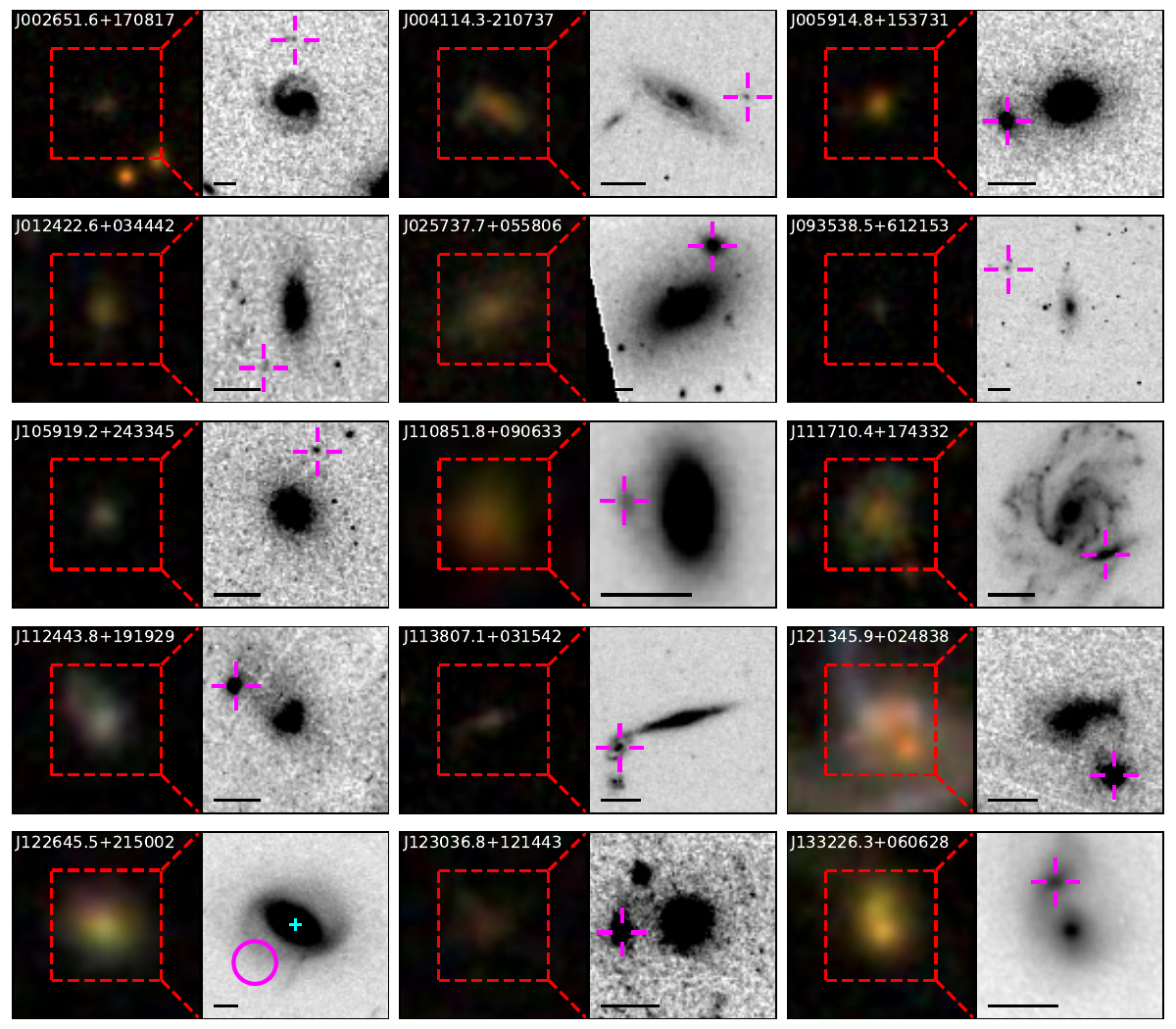}
\caption{\footnotesize{The final sample of \Samp~HLX candidates with \hst~imaging. The left panels show the SDSS $g$\,$+$\,$r$\,$+$\,$i$ composite color images. The right panels show the adopted \hst~image (Section \ref{sec:HST}). The crosshairs denote the centroid position of the HLX candidate \hst~counterpart detections (Section \ref{sec:Detect}). If no counterpart is detected, the X-ray source position is circled. In each panel, North is up and East is to the left, the scale bar is 2$''$, and the HLX candidate names are labeled. Host galaxy extended features connecting to the HLX \hst~counterpart are seen in J0026, J0059, J0257, J1108, J1124, J1213, J1230, J1332, J1400, J1429, J1449, and J2336, and they are listed in Table \ref{tab:Samp}.}}
\label{fig:examples}
\end{figure*}

\begin{figure*}[ht!]
\ContinuedFloat
\includegraphics[width=0.96\textwidth]{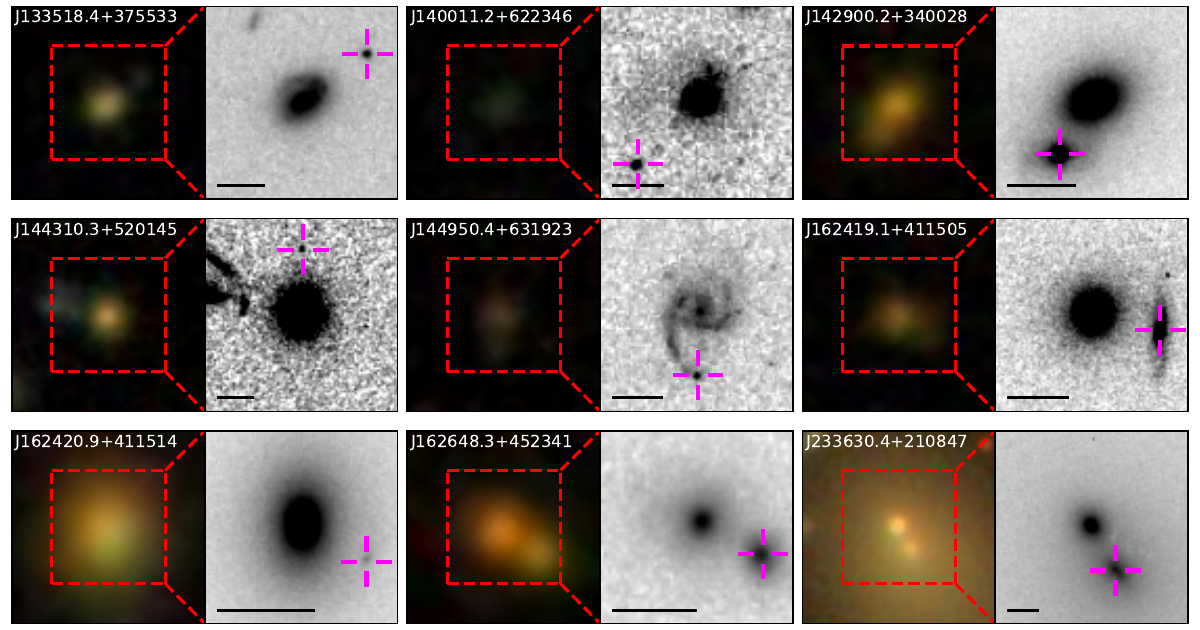}
\caption{\footnotesize{Continued.}}
\label{fig:examples}
\end{figure*}
 
\section{Procedure}
\label{sec:proc}

Here we describe our procedure for building a sample of HLX candidates with stellar core detections or upper limits using archival \hst~imaging. These steps consist of selecting the initial sample of HLX candidates (Section \ref{sec:XRay}), developing the subsample with \hst~imaging (Section \ref{sec:HST}), detecting \hst~counterparts to the HLXs (Section \ref{sec:Detect}), and measuring their stellar masses (Section \ref{sec:StellarMasses}).  \\ \\

\subsection{HLX Candidate Selection and Contamination Fraction Estimates}
\label{sec:XRay}

The sample of HLX candidates is drawn from \citet{Barrows:2019} and based on the procedure from \citet{Barrows:2016}. In short, these steps consisted of first matching X-ray sources from the CSC Version 2 with host galaxies from the SDSS photometric catalog (within two Petrosian radii), registering their images, and determining the best host galaxy redshifts. Active galactic nuclei (AGN) were then selected as X-ray sources with 0.5$-$7\,keV rest-frame (unabsorbed) luminosities (\LXB) in the range \LXB\,$=$\,$10^{41-42}$\,\uLum~and greater than the expected contribution from XRBs and hot interstellar medium gas (based on the host galaxy star formation rates and stellar masses) by $\ge$\,5 times the uncertainty. Finally, HLX candidates were selected as X-ray AGN that are spatially offset from the host galaxy centroid by $\ge$\,5 times the relative offset uncertainty (quadrature sum of the uncertainties in the astrometric solution, the X-ray source centroid, and the host galaxy centroid) in either of the right ascension or declination dimensions. The sample used in this work has been expanded by omiting the restriction of \LXB\,$\le$\,$10^{42}$\,\uLum~to not introduce biases in black hole mass or Eddington ratio. Moreover, the values of \LXB~have been remeasured based on the procedure from \citet{Barrows:2022}, and host galaxy star formation rates and stellar masses have been updated based on broadband spectral energy distribution (SED) modeling following \citet{Barrows:2022}.

We estimate the expected fraction of the HLXs that are unrelated (i.e., background or foreground) sources following the procedure from \citet{Barrows:2022}, which is based on the methodology applied to previous samples of off-nuclear X-ray sources \citep[e.g.,][]{Walton:2011,Sutton:2012,Zolotukhin:2016,Earnshaw:2019,Walton:2022,Barrows:2022}. In short, this procedure consists of first computing the number of X-ray sources expected to randomly be within the solid angle of each galaxy in which we search for X-ray sources (i.e., a circle of two Petrosian radii, after excising the nuclear region defined by the offset threshold). We then determine the expected number of X-ray sources in this area using the resolved 0.5$-$7\,keV point source density function of \citet{Masini:2020} and the effective limiting sensitivities at each galaxy position. We set the effective limiting sensitivity to the flux corresponding to an observed 0.5$-$7\,keV luminosity of $10^{41}\,$erg s$^{-1}$ (adopted HLX lower limit) at the host galaxy redshift, or otherwise the 0.5$-$7\,keV CSC limiting sensitivity at the galaxy position if it is larger. The estimated contamination fraction among this initial sample (total number of expected X-ray sources divided by the total number of HLX candidates) is $40^{+6}_{-10}$\%
(qualitatively similar to other samples of off-nuclear X-ray sources; e.g., \citealt{Zolotukhin:2016,Gong:2016,Earnshaw:2019,Walton:2022}).

From a spatial match with the NASA Extragalactic Database, the sample is cleansed of known contaminating background or foreground sources (based on redshifts) and of known jets or gravitational lenses. We also remove X-ray sources with spatial extents greater than the point spread function (PSF) $1\sigma$ upper bound and those with spatial extents overlapping with one or more other X-ray sources.

\begin{deluxetable*}{lcccccccccc}
\tabletypesize{\footnotesize}
\tablecolumns{11}
\tablecaption{HLX Candidates with \hst~Imaging.}
\tablehead{
\colhead{CSC Name} &
\colhead{$z$} &
\colhead{log$\left(\frac{L_{X}}{\rm{erg\,s^{-1}}}\right)$} &
\colhead{log$\left(\frac{M_{\rm{\star,Host}}}{M_{\odot}}\right)$} &
\colhead{ID$_{\rm{Host}}$} &
\colhead{$\frac{\Delta S}{\rm{kpc}}$} &
\colhead{$m_{\rm{HLX}}$} &
\colhead{log$\left(\frac{M_{\rm{\star,HLX}}}{M_{\odot}}\right)$} &
\colhead{log$\left(\frac{F_{X}}{F_{O}}\right)$} &
\colhead{Int} &
\colhead{Filter} \\
\colhead{1} &
\colhead{2} &
\colhead{3} &
\colhead{4} &
\colhead{5} &
\colhead{6} &
\colhead{7} &
\colhead{8} &
\colhead{9} &
\colhead{10} &
\colhead{11}
}
\startdata
J002651.6+170817 & 0.185$^a$ & $42.00_{-0.45}^{+0.48}$ & $10.19_{-0.45}^{+0.41}$ & 24 & $19.29\pm1.52$ & 28.05 & $7.01_{-0.30}^{+0.19}$ & $2.10_{-0.19}^{+0.12}$ & T/S & F814W \\ 
J004114.3$-$210737 & 0.270$^b$ & $43.06_{-0.78}^{+0.78}$ & $10.87_{-0.46}^{+0.42}$ & 18 & $12.34\pm2.03$ & 22.99 & $9.15_{-0.30}^{+0.19}$ & $0.98_{-0.08}^{+0.07}$ & N & F850LP \\ 
J005914.8+153731 & 0.366$^b$ & $43.06_{-0.56}^{+0.56}$ & $11.11_{-0.64}^{+0.56}$ & 11 & $14.83\pm2.50$ & 22.03 & $10.17_{-0.17}^{+0.14}$ & $0.25_{-0.14}^{+0.11}$ & T & F775W \\ 
J012422.6+034442 & 0.448$^b$ & $42.03_{-0.35}^{+0.38}$ & $10.60_{-0.56}^{+0.50}$ & 35 & $18.06\pm2.82$ & 28.29 & $7.54_{-0.22}^{+0.16}$ & $1.35_{-0.25}^{+0.15}$ & N & F606W \\ 
J025737.7+055806 & 0.149$^b$ & $41.71_{-0.51}^{+0.51}$ & $10.78_{-0.56}^{+0.50}$ & 37 & $12.43\pm1.27$ & 21.19 & $9.34_{-0.21}^{+0.16}$ & $-0.49_{-0.16}^{+0.12}$ & T & F160W \\ 
J093538.5+612153 & 0.184$^b$ & $41.59_{-0.49}^{+0.49}$ & $9.66_{-0.30}^{+0.28}$ & 34 & $23.17\pm1.51$ & 23.63 & $8.46_{-0.54}^{+0.24}$ & $-0.01_{-0.17}^{+0.12}$ & N & F814W \\ 
J105919.2+243345 & 0.161$^b$ & $41.54_{-0.46}^{+0.49}$ & $9.87_{-0.38}^{+0.35}$ & 35 & $8.23\pm1.36$ & 24.84 & $8.00_{-0.38}^{+0.21}$ & $0.52_{-0.18}^{+0.12}$ & N & F814W \\ 
J110851.8+090633 & 0.442$^b$ & $42.66_{-0.49}^{+0.49}$ & $10.98_{-0.63}^{+0.55}$ & 37 & $8.14\pm2.80$ & 24.85 & $8.63_{-0.17}^{+0.15}$ & $1.27_{-0.17}^{+0.12}$ & T & F140W \\ 
J111710.4+174332 & 0.350$^b$ & $43.10_{-0.96}^{+1.00}$ & $11.27_{-0.64}^{+0.56}$ & 36 & $12.27\pm2.42$ & 20.47 & $10.05_{-0.17}^{+0.14}$ & $-0.00_{-0.05}^{+0.04}$ & N & F814W \\ 
J112443.8+191929 & 0.166$^a$ & $41.97_{-0.36}^{+0.38}$ & $10.03_{-0.55}^{+0.49}$ & 35 & $9.01\pm1.39$ & 20.14 & $9.39_{-0.22}^{+0.16}$ & $-0.84_{-0.25}^{+0.15}$ & T & F160W \\ 
J113807.1+031542 & 0.199$^b$ & $41.93_{-0.61}^{+0.64}$ & $9.98_{-0.36}^{+0.33}$ & 17 & $12.19\pm1.61$ & 21.82 & $9.40_{-0.42}^{+0.22}$ & $-0.44_{-0.12}^{+0.09}$ & N & F814W \\ 
J121345.9+024838 & 0.073$^a$ & $41.30_{-0.46}^{+0.49}$ & $10.35_{-0.69}^{+0.59}$ & 18 & $4.26\pm0.68$ & 14.01 & $10.31_{-0.15}^{+0.14}$ & $-2.20_{-0.18}^{+0.12}$ & T & F222M \\ 
J122645.5+215002 & 0.195$^b$ & $41.09_{-0.42}^{+0.42}$ & $10.75_{-0.56}^{+0.50}$ & 17 & $6.67\pm1.59$ $^c$ & 27.10$^d$ & 5.90$^d$ & 0.63$^d$ & N & F814W \\ 
J123036.8+121443 & 0.553$^b$ & $43.96_{-1.27}^{+1.27}$ & $11.37_{-0.55}^{+0.49}$ & 37 & $14.55\pm3.16$ & 21.24 & $11.02_{-0.22}^{+0.16}$ & $0.44_{-0.02}^{+0.02}$ & T & F850LP \\ 
J133226.3+060628 & 0.207$^a$ & $41.66_{-0.41}^{+0.41}$ & $10.95_{-0.69}^{+0.59}$ & 27 & $4.54\pm1.66$ & 17.61 & $11.08_{-0.15}^{+0.14}$ & $-2.29_{-0.21}^{+0.14}$ & T & F105W \\ 
J133518.4+375533 & 0.310$^a$ & $42.50_{-0.45}^{+0.47}$ & $10.90_{-0.57}^{+0.50}$ & 37 & $16.16\pm2.24$ & 22.46 & $9.53_{-0.21}^{+0.16}$ & $-0.08_{-0.19}^{+0.13}$ & N & F814W \\ 
J140011.2+622346 & 0.310$^a$ & $42.03_{-0.43}^{+0.43}$ & $9.91_{-0.50}^{+0.46}$ & 15 & $16.66\pm2.24$ & 24.02 & $7.99_{-0.25}^{+0.17}$ & $0.38_{-0.20}^{+0.14}$ & T & F814W \\ 
J142900.2+340028 & 0.352$^a$ & $41.62_{-0.01}^{+0.04}$ & $11.42_{-0.67}^{+0.57}$ & 1 & $10.18\pm2.43$ & 20.17 & $10.82_{-0.16}^{+0.14}$ & $-1.94_{-1.00}^{+0.32}$ & T & F814W \\ 
J144310.3+520145 & 0.156$^b$ & $41.39_{-0.35}^{+0.37}$ & $10.58_{-0.59}^{+0.52}$ & 8 & $10.25\pm1.33$ & 24.29 & $8.25_{-0.20}^{+0.15}$ & $0.22_{-0.25}^{+0.15}$ & N & F702W \\ 
J144950.4+631923 & 0.341$^b$ & $42.07_{-0.53}^{+0.55}$ & $10.26_{-0.30}^{+0.28}$ & 8 & $13.15\pm2.38$ & 23.71 & $9.19_{-0.54}^{+0.24}$ & $-0.50_{-0.15}^{+0.11}$ & T/S & F606W \\ 
J162419.1+411505 & 0.284$^b$ & $43.07_{-0.80}^{+0.80}$ & $11.49_{-0.42}^{+0.38}$ & 27 & $10.25\pm2.10$ & 22.23 & $10.68_{-0.34}^{+0.20}$ & $0.47_{-0.07}^{+0.06}$ & N & F606W \\ 
J162420.9+411514 & 0.181$^a$ & $41.63_{-0.35}^{+0.35}$ & $11.21_{-0.54}^{+0.48}$ & 11 & $4.76\pm1.50$ & 25.38 & $8.42_{-0.23}^{+0.17}$ & $0.49_{-0.25}^{+0.16}$ & N & F606W \\ 
J162648.3+452341 & 0.465$^a$ & $42.75_{-0.55}^{+0.58}$ & $11.90_{-0.69}^{+0.59}$ & 13 & $10.30\pm2.88$ & 19.88 & $11.40_{-0.15}^{+0.14}$ & $-1.13_{-0.14}^{+0.10}$ & N & F785LP \\ 
J233630.4+210847 & 0.055$^a$ & $41.61_{-0.99}^{+1.02}$ & $11.99_{-0.59}^{+0.52}$ & 33 & $3.71\pm0.52$ & 20.75 & $10.21_{-0.20}^{+0.15}$ & $-1.11_{-0.05}^{+0.04}$ & D & F555W
\enddata
\tablecomments{Column 1: CSC source; column 2: host galaxy best redshift; column 3: rest-frame, 0.5$-$7\,keV, unabsorbed luminosity; column 4: host galaxy stellar mass; column 5: host galaxy best fit SDSS photometric redshift template ID (Section \ref{sec:StellarMasses}); column 6: projected physical offset of the HLX \hst~counterpart from the host galaxy centroid; columns 7$-$9: HLX \hst~counterpart AB magnitude, stellar mass, and X-ray (0.3$-$3.5\,keV) to optical ($V$-band) flux ratio; column 10: morphological feature indicating interaction between the HLX \hst~counterpart and the host galaxy (`T'=tidal feature, `S'=spiral feature, `D'=dust lane, and `N'=none); and column 11: adopted \hst~filter.\\ $^a$Spectroscopic redshift \\ $^{b}$Photometric redshift \\ $^c$When no \hst~counterpart is detected, the HLX candidate position is taken to be that of the X-ray source. \\ $^d$ Based on the upper limit \hst~detection.}
\label{tab:Samp}
\end{deluxetable*}

\begin{figure}[ht!]
\includegraphics[width=0.48\textwidth]{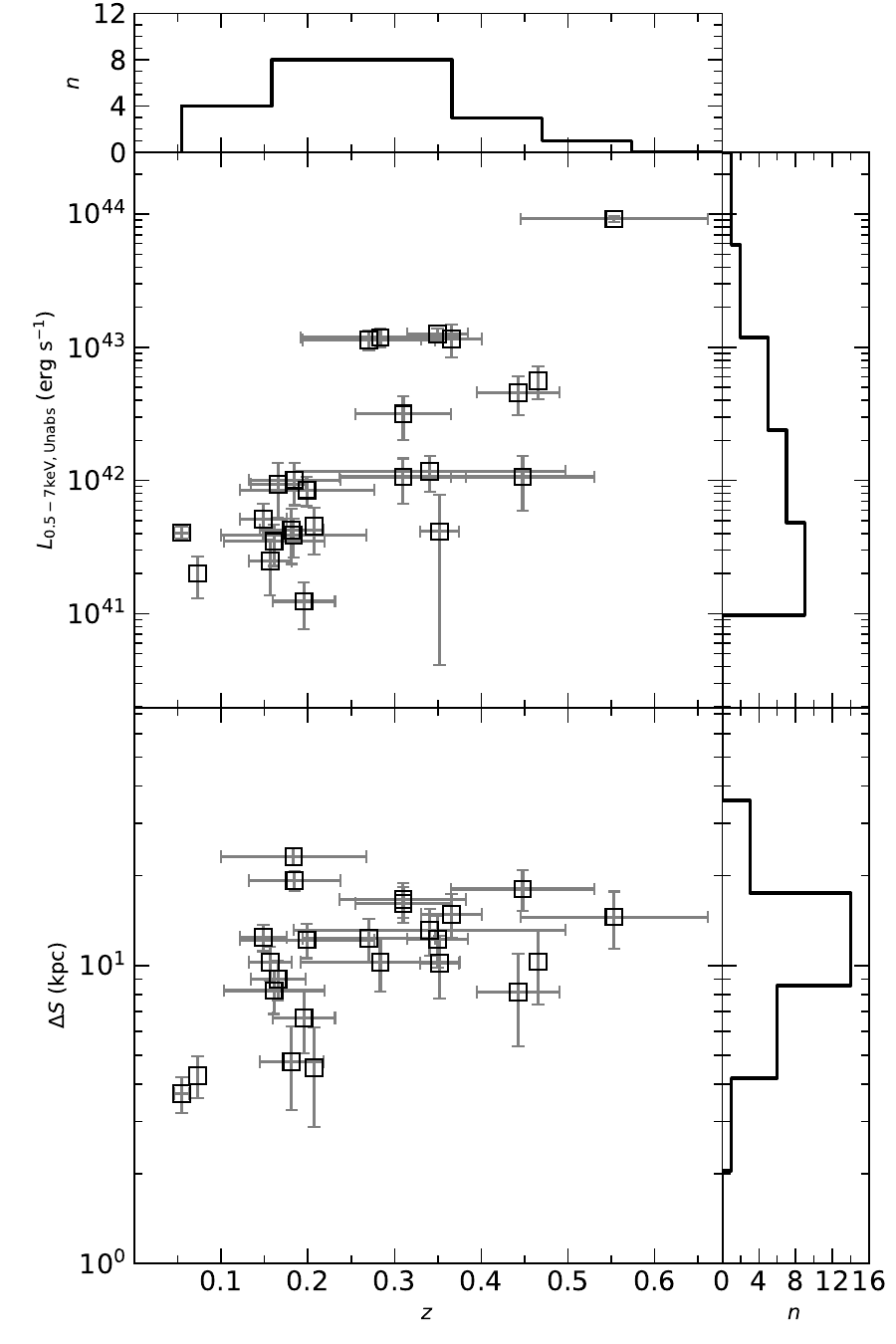}
\caption{\footnotesize{Rest-frame, unabsorbed 0.5$-$7\,keV luminosity (\LXB; top) and projected physical offset from the host galaxy centroid (\DeltaS; bottom) against host galaxy redshift ($z$) for the final sample of \Samp~HLX candidates with \hst~imaging (Section \ref{sec:HST}). The positive dependence of both \LXB~and \DeltaS~with $z$ reflect the CSC flux sensitivity and spatial resolution limits, respectively.}}
\label{fig:Z_DELTAS_LX}
\end{figure}

\subsection{HLX Candidates with HST Imaging}
\label{sec:HST}

We crossmatch the sample of HLX candidates from Section \ref{sec:XRay} with the \emph{Hubble}~Source Catalog Version 3 using a conservatively large match radius of 10'' (to account for uncertainties in the relative astrometry before registration with the \ch~images). If an HLX candidate is matched with \hst~images in multiple filters, we adopt the image in the filter with the reddest effective wavelength (for the optimal tracer of older stars in galaxy stellar bulges). We then register the \hst~images to the SDSS images (following the procedure described in \citealp{Comerford:2015} and \citealp{Barrows:2018}). Since the \ch~images are previously registered to the SDSS images (Section \ref{sec:XRay}), this puts them on a common celestial reference frame with the \hst~images. We then require that the \hst~images cover the host galaxy centroid and the HLX candidate position. If an image does not meet this requirement,  then that with the next reddest filter effective wavelength is adopted, if available.

This yields \Samp~HLX candidates with \hst~imaging (Figure \ref{fig:examples}). One (J2336) is a previously known HLX candidate \citep{Gong:2016} and another (J1213) is a previously known dual AGN \citep[i.e., the primary galaxy nucleus also hosts an AGN;][]{Liu:2011,Hou:2020,DeRosa:2023}. Of the HLX candidates from Section \ref{sec:XRay} that are removed as known contaminants, four of them pass the \hst~imaging selection criteria from this section. Since 0.40\,$\times$\,\SampwCont\,$=$\,11 contaminants are expected among the sample of \SampwCont~that includes these contaminants, 7 (11-4) contaminants are expected to remain in the final sample of \Samp~(\PercCorrContwErr\%).

The \hst~filter effective wavelengths (in the host galaxy rest-frame) have a mean value of 7520\,\AA~(the majority of flux at this wavelength is generally expected to come from the host galaxy; e.g., \citealt{Barrows:2021}). The adopted \hst~filters, along with the host galaxy redshifts and \LXB~values, are listed in Table \ref{tab:Samp}. For those with multiple \ch~observations, no significant flux variability is reported in the CSC. The distribution of \LXB~values (Figure \ref{fig:Z_DELTAS_LX}) shows that most of the HLX candidates are typical of low- to moderate-luminosity AGN. The positive dependence of \LXB~on redshift illustrates the inherent bias imposed by the CSC sensitivity limits on the initial selection (Section \ref{sec:XRay}).

\subsection{Detecting HST Counterparts}
\label{sec:Detect}

To search for \hst~detections associated with the HLX candidates from Section \ref{sec:HST}, we follow the procedure from \citet{Barrows:2017b} and \citet{Barrows:2018}. In brief, we first use \texttt{Source Extractor} \citep{Bertin:Arnouts:1996} to find significantly ($\ge$3$\sigma$) detected sources. We then fit each of these sources with a Sersic profile (using the \texttt{Source Extractor} source positions, fluxes, and spatial extents as initial inputs) with \texttt{Galfit} \citep{Peng:2010}, along with a globally measured background.  For each source, we test if it is spatially resolved by separately fitting a PSF model (constructed from the images) and testing if using the Sersic component instead improves the fit at the 99.73\%~level based on an $F$-test. If so,  we also test if the model is improved by adding a PSF component to the Sersic component (to account for potentially significant contribution from the X-ray point source). 

We find 23 detections out of the \Samp~HLX candidates in the final sample (they are indicated in Figure \ref{fig:examples}). The image model for each detection requires a Sersic component (i.e., can not be adequately fit with only a PSF component). In no cases is the addition of a PSF component to the Sersic component necessary based on our adopted statistical threshold. However, the fluxes modeled by the Sersic components may also include emission from point sources that can not be distinguished with the current imaging. This is in contrast to the well-studied source HLX-1 \citep[e.g.,][]{Farrell:2009} for which the optical counterpart is dominated by a PSF component \citep{Soria:2010}, though it is similar to other HLXs with optical counterpart stellar masses comparable to our sample \citep{Lin:2016,Lin:2020}. The \hst~counterpart magnitudes and their projected physical offsets from the host galaxy centroids (\DeltaS) are listed in Table \ref{tab:Samp}. The positive dependence of \DeltaS~with $z$ illustrates the inherent bias imposed by the \ch~spatial resolution on the initial selection (Section \ref{sec:XRay}).

The \hst~imaging depth allows for faint features to be detected, and the strongest HLX candidates are those for which the imaging reveals morphological evidence for interaction between the host galaxy and the stellar counterpart. This evidence (in the form of host galaxy tidal features, spiral arms, or dust lanes connecting to the HLX \hst~counterpart) are seen in 12 cases (Figure \ref{fig:examples} and Table \ref{tab:Samp}). While imaging in shorter wavelength filters is available for some of the HLX candidates, this imaging does not provide additional evidence for \hst~counterparts, interactions, or PSF components beyond those detected in the reddest usable filters.

\begin{figure*}[ht!]
\includegraphics[width=0.96\textwidth]{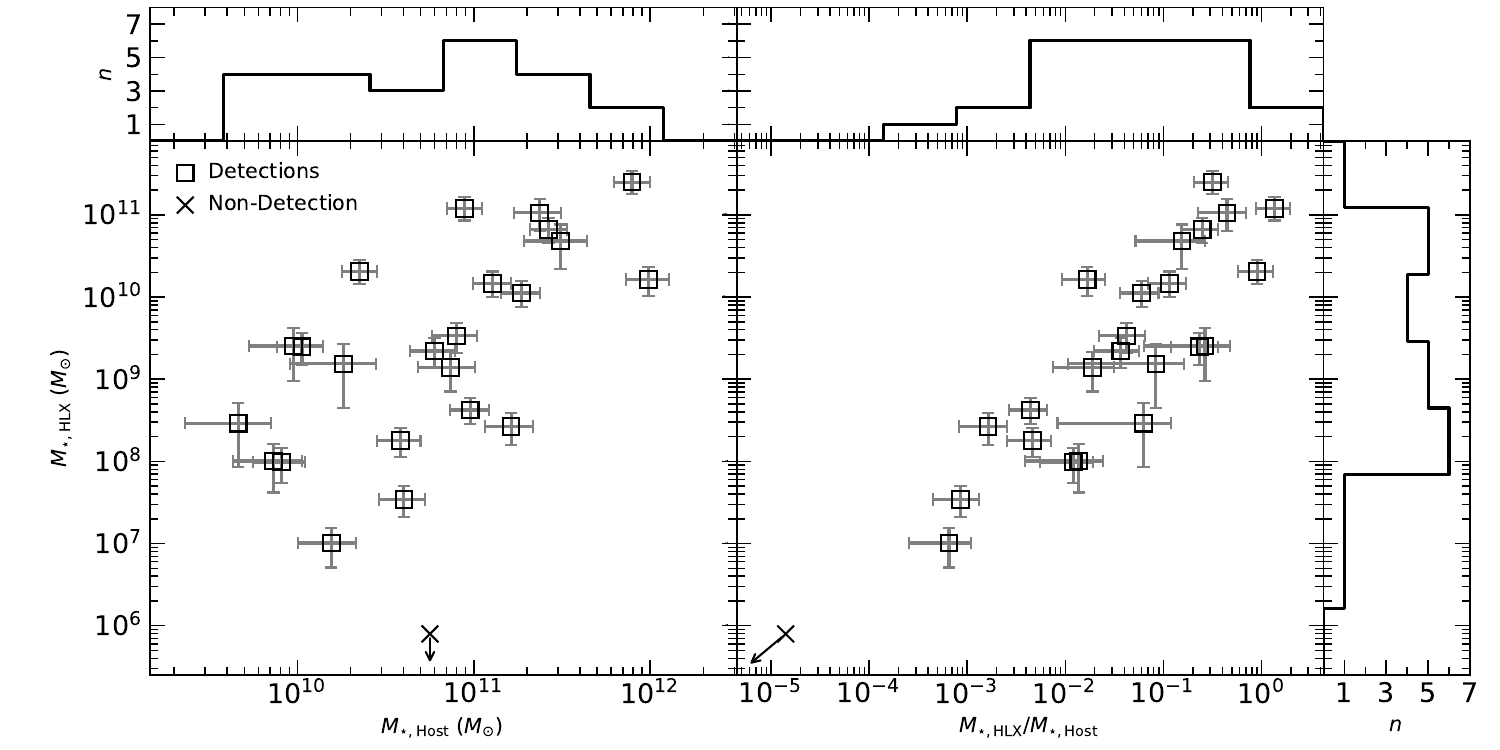}
\caption{\footnotesize{HLX candidate stellar core masses (\MstarTwo) against host galaxy stellar masses (\MstarOne; left) and ratios between \MstarTwo~and \MstarOne~(\MRATIO; right). The squares and the `x' denote \hst~counterpart detections and the upper limit, respectively. More massive stellar cores are generally associated with more massive host galaxies and with major mergers.}}
\label{fig:MSTAR_MSTAR_Prim_MRATIO}
\end{figure*}

\subsection{Stellar Masses}
\label{sec:StellarMasses}

Stellar masses of the HLX candidate \hst~counterparts and their host galaxies are computed by applying their flux ratios (from the \hst~image detections; Section \ref{sec:Detect}) to the total stellar masses (Section \ref{sec:XRay}; assuming they represent the combined stellar mass of the host and HLX \hst~counterpart). While this procedure also assumes the HLX stellar cores have the same SEDs (and hence mass-to-light ratios; $M/L$) as their host galaxies, individually constraining stellar populations for each counterpart in the sample is not possible. The host galaxy best-fit SED template IDs (based on the SDSS photometric redshift pipeline\footnote{https://www.sdss4.org/dr17/algorithms/photo-z/}; \citealp{Dobos:2012}), along with the stellar masses of the host galaxies and HLX \hst~counterparts are listed in Table \ref{tab:Samp}.

The collection of \hst~imaging we use has a heterogeneous array of depths. To account for this, the minimum detectable stellar counterpart mass for each image is computed by converting the 3$\sigma$ global background value (from the image modeling; Section \ref{sec:Detect}) to a stellar mass, again assuming the same $M/L$ value as for the host galaxy. This value is used as the upper limit for the HLX candidate without an \hst~counterpart detection.

Figure \ref{fig:MSTAR_MSTAR_Prim_MRATIO} shows the distribution of HLX candidate stellar core masses (\MstarTwo) against the host galaxy stellar masses (\MstarOne) and their ratios (\MRATIO). The median value of \MstarTwo~is \MSTARMedian\,\Msun~(near the conventional dwarf galaxy stellar mass threshold of $10^{9.5}$\,\Msun). The values extend upward to $\sim$\,$10^{11}$\,\Msun~and downward to low-mass stellar cores ($\sim$\,$10^7$\,\Msun). Despite the deep \hst~imaging (down to stellar mass sensitivities of $\sim$\,$10^5$\,\Msun), no detected counterparts have stellar masses below $10^{7}$\,\Msun.

The stellar masses are all larger than those of the most massive Milky Way globular clusters observed \citep[e.g.,][]{Baumgardt:2018}. While some globular clusters are known to have masses of a few $10^6$\,\Msun, they are likely the remnants of accreted dwarf galaxy satellites \citep[e.g.,][]{Bekki:2003,Forbes:2010} rather than from in-situ formation. Hence, the stellar core masses of our HLX sample suggest they originate from the nuclei of external galaxies that have merged, or are merging, with the host galaxy. For two HLX candidates (J0026 and J1449), the host galaxy features connecting to the HLX \hst~counterpart have a morphological resemblance to spiral arms (and the hosts appear to be late-type). Since the \hst~counterparts may instead be regions of star formation, in these cases the possibility of exceptionally luminous XRBs or supernovae is considered (Section \ref{sec:connection}).

The largest values of \MRATIO~are near unity, consistent with traditional major galaxy mergers. As shown in Figure \ref{fig:examples}, these systems are visually similar to the types of mergers often identified morphologically based on clear visual disturbances. However, the dominance of \MRATIO~values significantly less than unity among the sample likely reflects the selection of X-ray sources that are not at the photometric centers of SDSS galaxies, resulting in the host galaxy contributing most of the flux.  Due to the insensitivity of the original selection to morphology, the \MRATIO~distribution spans more than three orders of magnitude (down to values of $\sim$\,$10^{-3}$ for HLXs with detections).

\begin{figure*}[ht!]
\includegraphics[width=\textwidth]{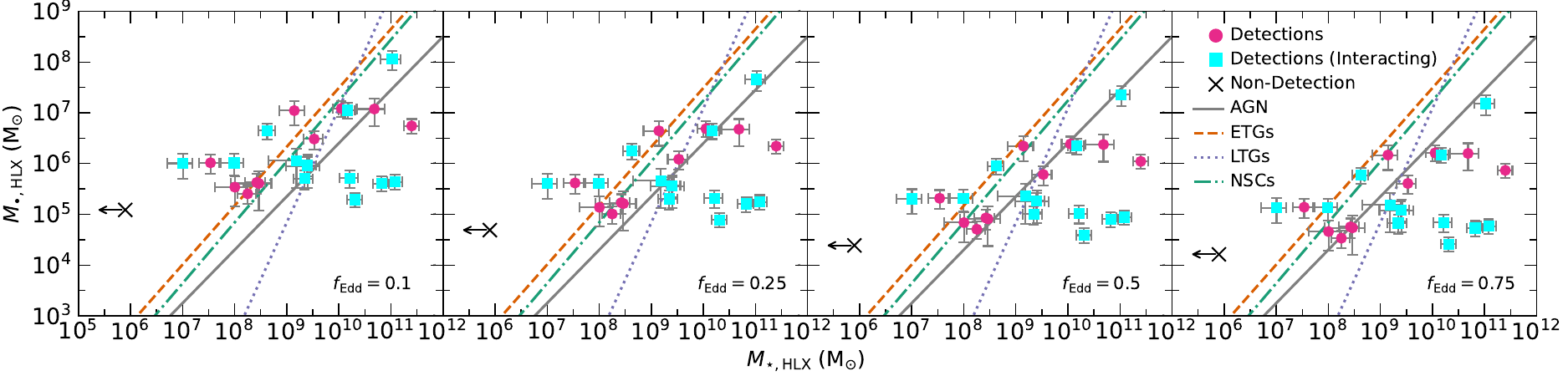}
\caption{\footnotesize{Black hole mass (\MBHHLX) against stellar core mass (\MstarTwo) for our final sample of \Samp~HLX candidates. Values of \MBHHLX~are estimated from \LXB, the X-ray bolometric correction function of \citet{Duras:2020}, and assuming Eddington ratios (\fEdd) of (from left to right) 0.1, 0.25, 0.5, and 0.75.  The solid (gray), dashed (orange), dotted (purple), and dash-dotted (green) lines indicate the expected host galaxy and black hole mass scaling relations for AGN host galaxies (\citealp{Reines:2015}), early- and late-type quiescent galaxies (E/LTGs; \citealp{Graham:2023}), and nuclear star clusters (NSCs; \citealp{Scott:2013}), respectively. \hst~counterpart detections with and without morphological evidence for interaction with their host galaxies are denoted by cyan squares and magenta circles, respectively. The source without a detection is denoted by an `x'. Assuming statistical agreement with the relations, the average best-fit \fEdd~values are \FEddFitScalRelaAGN, \FEddFitScalRelaETG, \FEddFitScalRelaLTG, and \FEddFitScalRelaNSC~for AGN, ETGs/LTGs, and NSCs, respectively.}}
\label{fig:MBH_MSTAR}
\end{figure*}

\begin{figure}[ht!]
\includegraphics[width=0.48\textwidth]{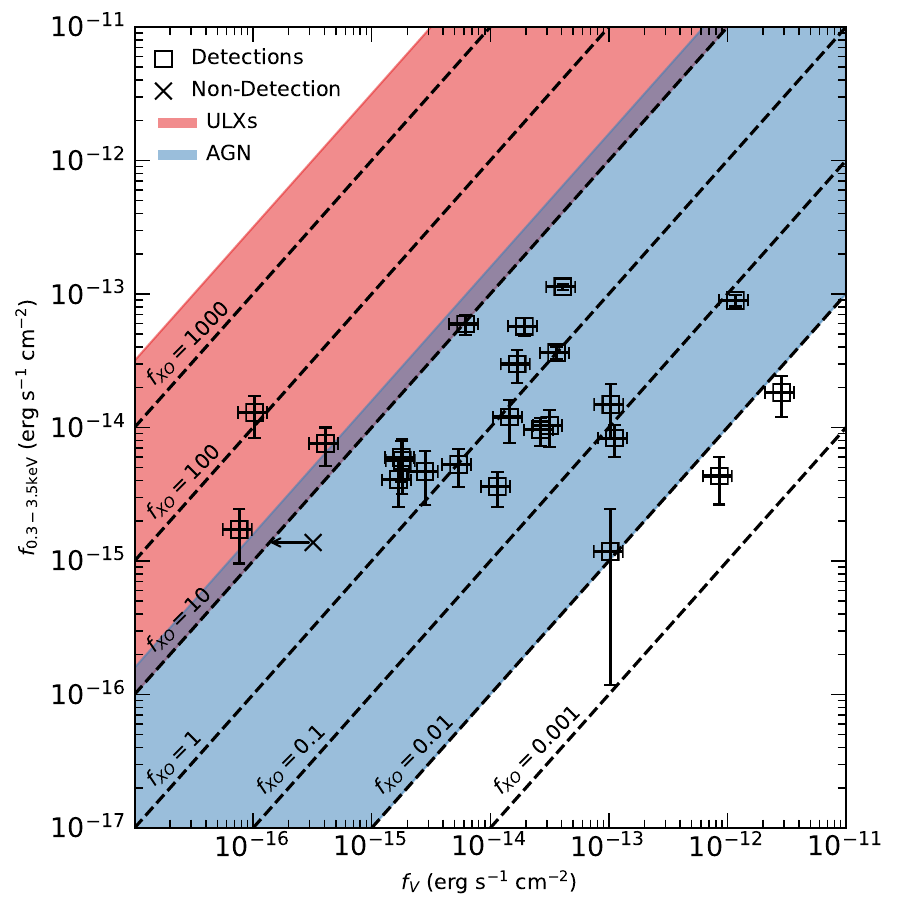}
\caption{\footnotesize{X-ray (0.3$-$3.5\,keV) against optical ($V$-band) fluxes for the full sample of \Samp~HLX candidates. The detections and non-detection are denoted by squares and an `x', respectively. X-ray-to-optical ratios ($f_{XO}$\,$=$\,$f_{\rm{0.3-3.5keV}}/f_{V}$) from 0.001 to 1000, in steps of one dex, are shown as dashed lines. The shaded regions represent the typical values for AGN (blue; \citealp[e.g.,][]{Stocke:1991,Lin:2012}) and ULXs (red; using the 90\%~lower bound of the sample from \citealp{Tao:2011}). The majority of the sample is consistent with the upper end of $f_{XO}$ values typically seen among AGN, while a few have significantly larger values typically only seen among lower mass accretors.}}
\label{fig:FX_FO}
\end{figure}

\section{Nature of the HLXs}
\label{sec:nature}

In this section we explore possible physical scenarios for the HLX candidates. We first examine the host galaxy and black hole mass scaling relations that describe the sample (Section \ref{sec:IMBH_Merge}). We then derive constraints on the mass distributions of the black holes powering the HLXs (Section \ref{sec:Masses}). Finally, we consider alternative scenarios beyond that of external galaxy mergers for individual sources (Section \ref{sec:Alt}).

\subsection{Massive Black Holes from Galaxy Mergers}
\label{sec:IMBH_Merge}

If HLXs are primarily the result of galaxy mergers, the masses of the stellar cores they reside in can be used to derive estimates of their black hole masses (\MBHHLX) by assuming a mass scaling relation between the original HLX host galaxy stellar core and their nuclear black holes. Since the previous evolutionary histories of the HLXs in our sample are unknown, in Figure \ref{fig:MBH_MSTAR} we compare to four different mass scaling relationships that represent those of AGN \citep{Reines:2015}, quiescent early- and late-type galaxies (E/LTGs; \citealp{Graham:2023}), and black holes within nuclear star clusters (NSCs; \citealp{Scott:2013}). The black hole masses for our sample are estimated by assuming an Eddington ratio (\fEdd; ratio of bolometric luminosity to Eddington luminosity, where \LEdd[\uLum]\,$=$\,$1.3\times10^{38}$\,\MBH[\Msun] and bolometric luminosities are computed from the HLX X-ray luminosities using the relation from \citealt{Duras:2020}). Figure \ref{fig:MBH_MSTAR} shows our sample in the black hole mass - stellar mass plane for Eddington ratios between 0.1 and 0.75. We solve for the Eddington ratios for which our sample best fits each relation based on the reduced $\chi^2$ statistic. The minimized reduced $\chi^2$ values are \RedChiSqFitScalRelaAGN, \RedChiSqFitScalRelaETG, \RedChiSqFitScalRelaLTG, and \RedChiSqFitScalRelaNSC~for the AGN, ETG, LTG, and NSC scaling relations, respectively, and the corresponding Eddington ratios are \FEddFitScalRelaAGN, \FEddFitScalRelaETG, \FEddFitScalRelaLTG, and \FEddFitScalRelaNSC.

We note that the average Eddington ratio of our sample will be overestimated if the X-ray bolometric corrections are smaller than those of the typical AGN population. As shown in Figure \ref{fig:FX_FO}, while the majority of HLXs in our sample have X-ray-to-optical ratios\footnote{\FX~is the 0.3$-$3.5\,keV observed flux and \FO~is the $V$-band observed flux \citep{Maccacaro:1988}. The X-ray spectral models (Section \ref{sec:XRay}) are used to calculate the observed 0.3$-$3.5\,keV flux, and the SDSS $r$- and $i$-band stellar counterpart magnitudes are converted to $V$-band magnitudes using the conversion from \citet{Jester05}. } (\FXO) that are consistent with those of AGN ($\sim$\,0.01$-$10; e.g., \citealp{Maccacaro:1988,Stocke:1991,Lin:2012}), the \hst~imaging yields detections with \FXO~values up to $\sim$100 that are similar to nearby well-studied HLXs and strong IMBH candidates \citep[e.g.,][]{Webb:2014,Zolotukhin:2016,Tranin:2023} and which may correspond to smaller X-ray bolometric corrections. Furthermore, while the contribution of accretion disk emission to the optical flux is not possible to quantify given the limitations of the image modeling (Section \ref{sec:Detect}), it is likely to be smaller than in Figure \ref{fig:FX_FO} and may correspond to even smaller X-ray bolometric corrections.

\begin{figure}[t!]
\includegraphics[width=0.48\textwidth]{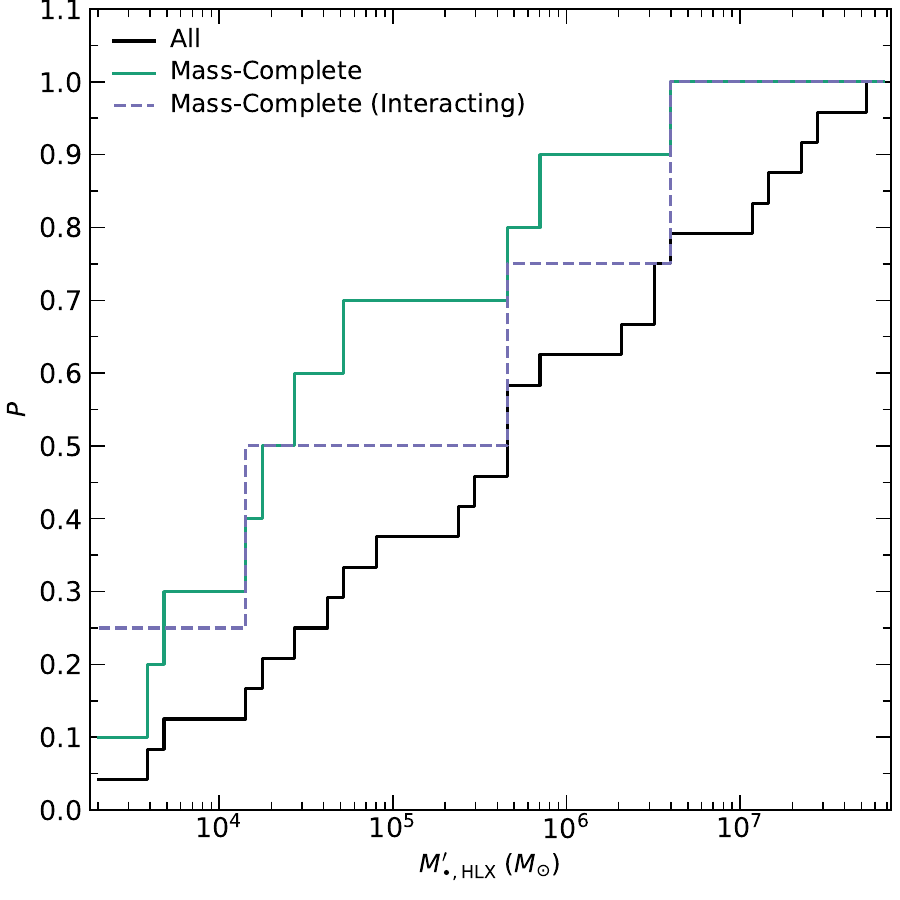}
\caption{\footnotesize{Cumulative sum (normalized to unity) of black hole mass lower limit estimates (\MBHHLXLo) based on the \Samp~HLX candidates from our sample and using the AGN mass scaling relation from \citet{Reines:2015}. The full sample is denoted by the solid black line. The mass-complete subsample (down to lower black hole mass limits of $10^3$\,\Msun) and the subset of it with morphological evidence for interaction with their host galaxies are denoted by the solid (green) and dashed (purple) lines, respectively. These lower limits reach down to the regime of black hole masses produced by primordial collapse of pristine gas clouds (\MBH\,$=$\,$10^{4-5}$\,\Msun) and runaway collisions in dense stellar clusters (\MBH\,$=$\,$10^{2-4}$\,\Msun).}}
\label{fig:MBH}
\end{figure} 

\subsection{Constraints on the Black Hole Mass Distribution of HLXs}
\label{sec:Masses} 

Since tidal stripping or threshing may have occured to the HLX stellar cores, the black hole masses derived from the mass scaling relations will be lower limits (\MBHHLXLo). Figure \ref{fig:MBH} shows the cumulative distribution of \MBHHLXLo~assuming the HLXs in our sample originally followed the AGN scaling relation of \citet{Reines:2015}, which provides the best overall fit to our sample (Section \ref{sec:IMBH_Merge}). After correcting for the estimated contamination fraction of \PercCorrCont\% (Section \ref{sec:HST}), the mass-complete subsample (complete down to lower black hole mass limits of $10^3$\,\Msun) suggests that up to \IMBHPercCorrAGN\% our sample may have black hole masses $<$\,$10^6$\,\Msun~(i.e., IMBHs). We find a consistent result (\IMBHPercIntAGN\%) when limiting to the strongest HLX candidates from our sample (those showing apparent interaction features with their host galaxy; Section \ref{sec:Detect}) and without statistically correcting for contamination. For comparison, when using the mass scaling relations for E/LTGs and NSCs, the overall IMBH fractions among our sample become \IMBHPercCorrETG\%, \IMBHPercCorrLTG\%, and \IMBHPercCorrNSC\%, respectively.

The most massive black hole seeds are predicted to form from the gravitational collapse of pristine primordial gas clouds (\MBH\,$\sim$\,$10^{4-5}$\,\Msun; e.g., \citealp{Loeb:1994,Lodato:2006}), while runaway collisions of stars in dense stellar clusters can form a black hole with a mass of \MBH\,$\sim$\,$10^{2-4}$\,\Msun~\citep[e.g.,][]{Devecchi:2009,DiCarlo:2021}. When correcting for contamination, we find up to \CollapsePercCorrAGN\%~of our mass-complete subsample has \MBHHLXLo~values consistent with the collapse scenario, and up to \ClusterPercCorrAGN\%~with formation in dense stellar clusters (assuming the AGN mass scaling relation). By implication, these HLXs may be probing a population of massive black hole seeds that have been accreted onto galaxy halos. However, also implied is that the majority have likely grown since their initial formation (further discussed in Section \ref{sec:connection}). Indeed, the luminosity threshold of $10^{41}$\,\uLum~imposes a lower black hole mass limit (per Eddington ratio), so HLXs may represent the massive end of the IMBH distribution. We note that the predicted lower black hole mass limits will be larger if the evolution of the overall sample was dominated by a mass scaling relation with a shallower slope at low black hole masses \citep[e.g.,][]{Mezcua:2017,Martin-Navarro:2018,Pacucci:2018} or if some of the black holes evolved to be overmassive when compared to the scaling relations \citep[e.g.,][]{Mezcua:2023,Pacucci:2023}.\\ \\

\begin{figure}[ht!]
\includegraphics[width=0.48\textwidth]{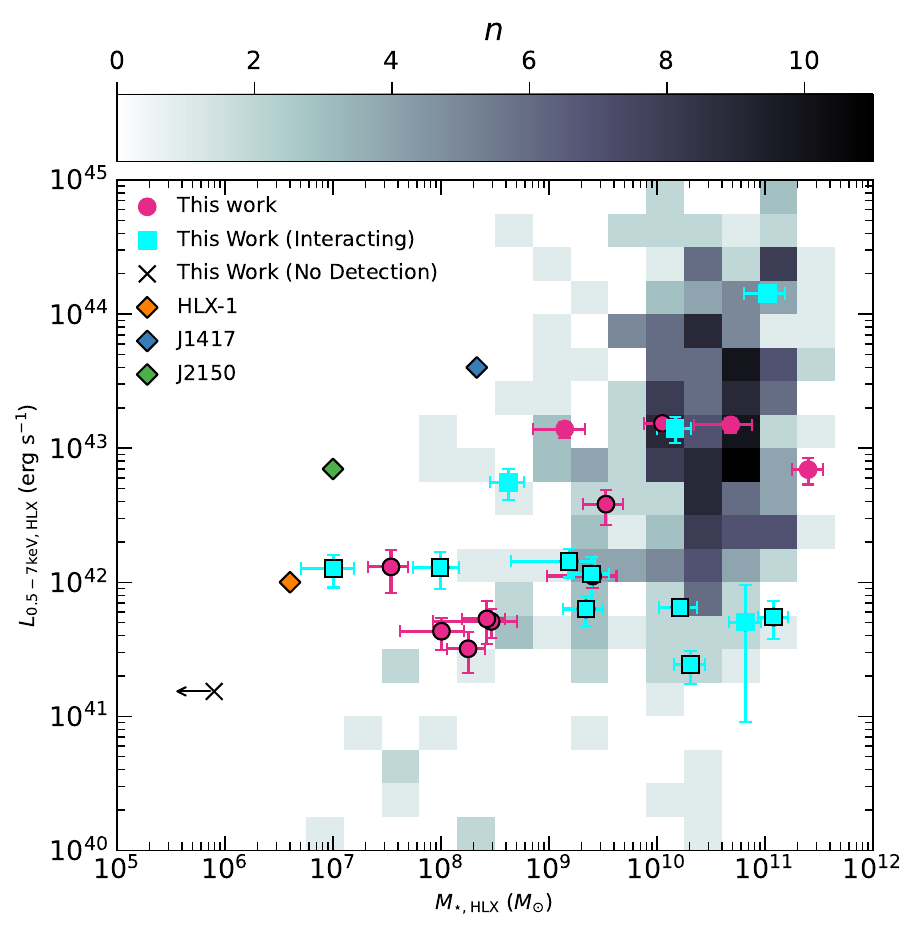}
\caption{\footnotesize{Rest-frame, 0.5$-$7\,keV, unabsorbed luminosity (\LXB) against stellar core mass (\MstarTwo) for our final sample of \Samp~HLX candidates. \hst~counterpart detections with and without morphological evidence for interaction with their host galaxies are denoted by cyan squares and magenta circles, respectively. The one without a detection is denoted by the `x'. We also show a sample of central X-ray AGN from the \ch~Deep Field South (CDFS; from \citealt{Magliocchetti:2020}). The subset of our sample that are complete down to the CDFS sample mass limit (black outlined symbols) have stellar core masses that are systematically lower, per unit luminosity, than those for the host galaxies of central AGN. For comparison, we also show confirmed HLXs with stellar core mass measurements from the literature (maximum X-ray luminosities plotted): 2XMM J011028.1$-$460421 \citep[HLX-1;][]{Farrell:2009}, 3XMM J141711.1$+$522541 \citep[mean mass estimate shown;][]{Lin:2016}, and 3XMM J215022.4$-$055108 \citep{Lin:2020}. The lowest masses of our sample are comparable to those of the confirmed HLXs, suggesting a similar physical origin.}}
\label{fig:LX_MSTAR}
\end{figure}

\subsection{Alternative Scenarios}
\label{sec:Alt}

For the HLX candidate with no \hst~counterpart detection (J1226; see Section \ref{sec:Detect}), if it is not a contaminant, the stellar mass upper limit of $\sim$\,$10^6$\,\Msun~would require significant luminosity enhancements and/or tidal stripping for the accreted satellite galaxy scenario. If the source is an XRB, its luminosity of \LXB\,$\approx$\,$2\times10^{41}$\,\uLum~would significantly exceed the prediction based on the normal galaxy population \citep[e.g.,][]{Lehmer:2010} and make it one of the most luminous ULXs known \citep[i.e.,][]{Israel:2017}. The XRB possibility is also relevant for the two HLXs that may reside in spiral arms (J0026 and J1449; see Section \ref{sec:StellarMasses}), though those sources are an order of magnitude even more luminous.

None of these three sources show AGN signatures based on mid-infrared colors (from the \wisetitle; \citealp{Wright:2010}, and the 75\%~reliability criterion from \citealt{Assef:2018}) or from radio detections (in the \vlasstitle~Epoch 1 Quick Look catalog; \citealp{Gordon:2020}), and none are coincident with known supernovae. An alternative possibility for these HLXs is IMBHs that formed in currently undetected globular clusters within their host galaxies \citep[e.g.,][]{Silk:1975,Haggard:2013,Wrobel:2015}. Hence, up to 3/\Samp~($\sim$\,12\%) of non-contaminant HLXs selected through our procedure may be XRBs or IMBHs that formed in-situ rather than originating from a merger.

Lastly, HLXs can potentially represent recoiling SMBHs following a galaxy merger and subsequent SMBH binary coalescence and asymmetric gravitational wave emission. In this scenario, theory predicts that the recoiling SMBH will carry a hypercompact stellar cluster of roughly the black hole mass \citep[e.g.,][]{Forbes:2008,Merritt:2009}. However, the non-detection of significant PSF components (Section \ref{sec:Detect}) may argue against the presence of hypercompact stellar clusters small enough to be consistent with hosting recoiling black holes. The possibility of this scenario can be further constrained with follow-up integral field spectroscopy to search for broad emission lines and motion relative to the host galaxy.

\section{The Role of Galaxy Mergers for IMBH Growth}
\label{sec:connection}

The deep \hst~imaging assembled in this study allows us to test if the galaxy mergers that produce HLXs also drive enhanced massive black hole growth by examining the connection between HLX luminosity and stellar core mass. Figure \ref{fig:LX_MSTAR} illustrates the positive relationship between \Mstar~and $L_{X}$ for our sample and for a sample of central (i.e., located in galaxy nuclei) X-ray AGN in the \ch~Deep Field South (CDFS) from \citet{Magliocchetti:2020}. This correlation reflects the intrinsic connection between nuclear massive black holes and galaxy bulges \citep[e.g.,][]{McLure:2002,Haring:2004,McConnell:2013}. However, when our sample is limited to the subset that is complete down to the same stellar mass limit as the CDFS sample (\Mstar\,$=$\,$10^7$\,\Msun), for a given X-ray luminosity the HLX stellar core masses are systematically lower relative to the host galaxy stellar masses of central AGN (median offset of \MedianLXMstarOffsetDex~dex). This result is qualitatively similar when limited to the subset showing morphological evidence for interaction with their host galaxies.

Among the HLXs with high-mass stellar cores (\MstarTwo\,$>$\,$10^{9.5}$\,\Msun), no significant offset from the comparison CDFS sample is seen. This result is consistent with numerical work suggesting the luminosity enhancements of AGN in major mergers are primarily determined by the host galaxy stellar mass, rather than merger-driven enhancements of the Eddington ratios \citep[e.g.,][]{Steinborn:2016,Weigel:2018}. Instead, the measured offset is driven by the HLXs with low-mass stellar cores. Moreover, Figure \ref{fig:LX_MSTAR} shows that the lowest mass HLX stellar cores in our sample are comparable to those of other confirmed HLXs which show a qualitatively similar offset from the expected relation between X-ray luminosity and host galaxy stellar mass, suggesting they are part of a similar population.

Assuming these offsets are due to enhanced luminosities, satellite accretion events may have a significant impact on IMBH growth. If these events are common, they may represent a non-negligible mechanism in the build-up of black hole mass in the Universe \citep[e.g.,][]{Barrows:2023}. However, no evidence is seen for the mass-normalized HLX luminosities to evolve with \DeltaS, suggesting that these enhanced $L_{X}$ to \Mstar~ratios are not strongly dependent on the stage of orbital evolution probed by our sample (though such a correlation may be washed out by random variations in projection angle and orbital trajectory).

\begin{figure}[ht!]
\includegraphics[width=0.48\textwidth]{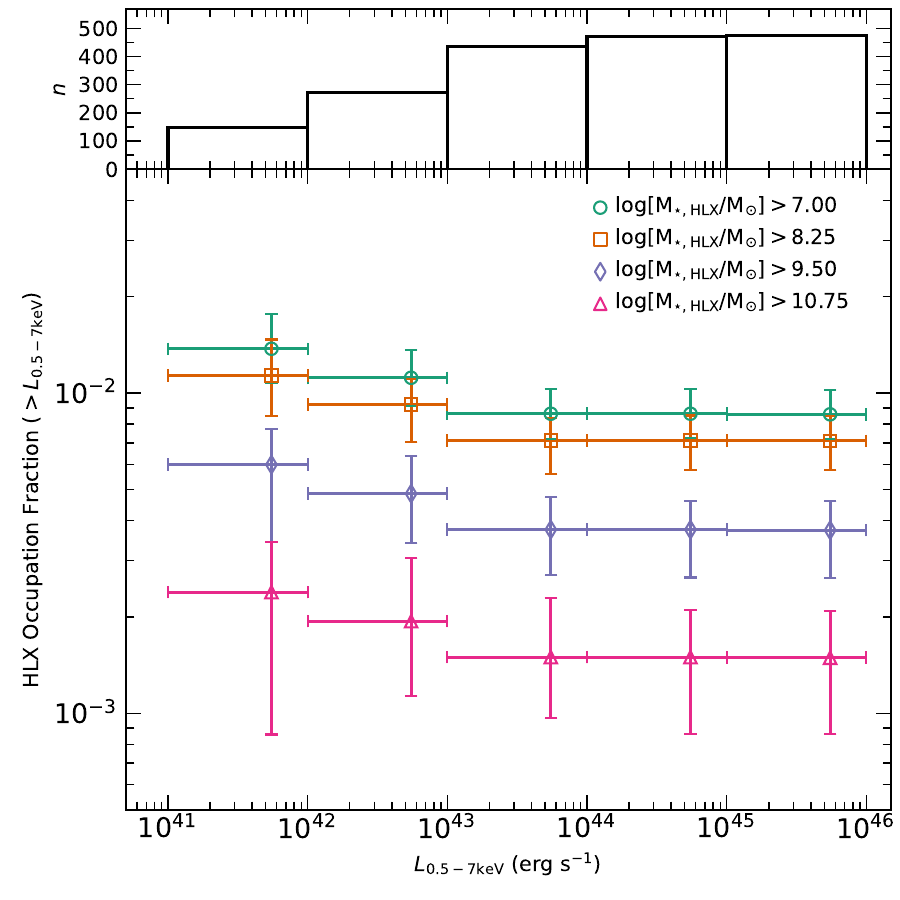}
\caption{\footnotesize{Number of HLXs (with the 0.5$-$7\,keV luminosity above a given limiting luminosity) per galaxy as a function of the limiting luminosity. Occupation fractions are shown for subsets with stellar core masses (\MstarTwo) above a given value (accounting for the limiting mass sensitivities). For reference, the number of galaxies in the parent sample in each luminosity bin are shown in the top panel. The occupation fraction declines nearly an order of magnitude over the full mass range of our sample, while the dependence on luminosity is significantly weaker.}}
\label{fig:Occ_Frac_MSTAR_LX}
\end{figure}

\begin{figure}[ht!]
\includegraphics[width=0.48\textwidth]{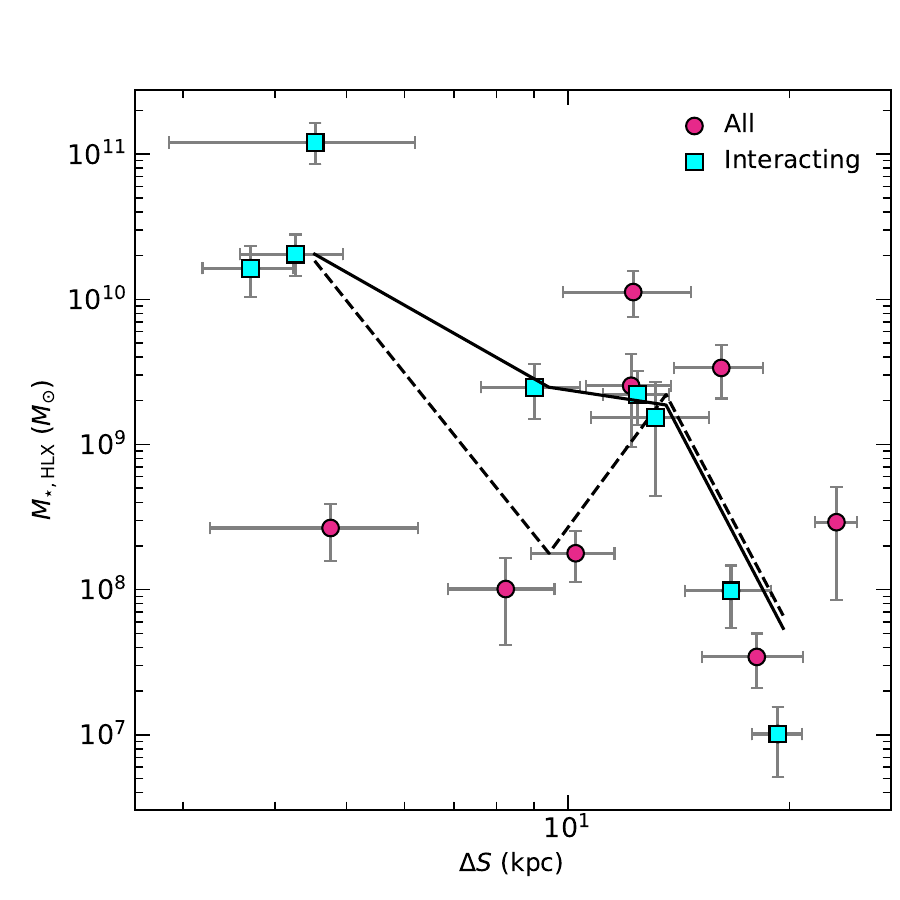}
\caption{\footnotesize{HLX stellar core mass (\MstarTwo) against projected physical offset from the host galaxy centroid (\DeltaS) for the subset of the final sample with HLX \hst~counterpart detections and that is complete down to \Mstar\,$=$\,$10^7$\,\Msun. Those with and without morphological evidence for interaction with their host galaxies are denoted by cyan squares and magenta circles, respectively. The dashed and solid lines indicate the median \MstarTwo~values in even logarithmic bins of \DeltaS~for all the data points and for the cyan squares, respectively. A negative correlation is seen (strongest for the cyan squares), suggesting that more massive stellar cores are more efficient at evolving toward the host galaxy nuclei through dynamical friction.}}
\label{fig:DELTAS_MSTAR_BIN}
\end{figure}

\begin{figure}[ht!]
\includegraphics[width=0.48\textwidth]{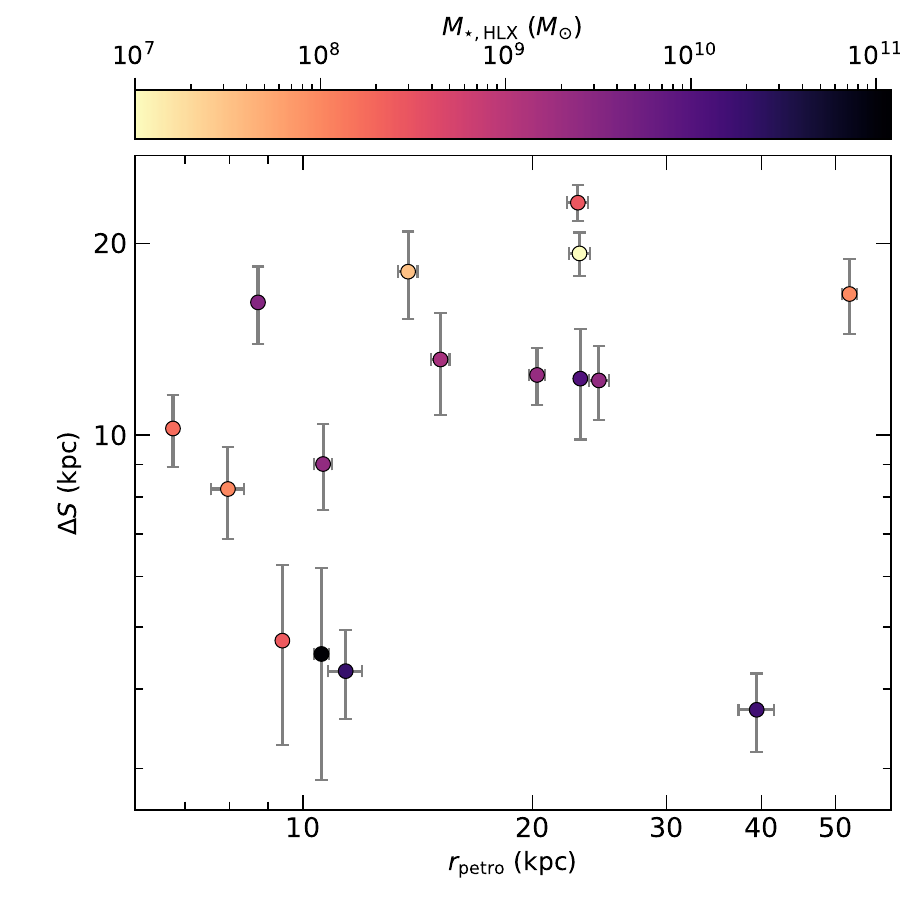}
\caption{\footnotesize{HLX stellar core projected nuclear physical offset (\DeltaS) against host galaxy characteristic physical size (Petrosian radius; $r_{\rm{petro}}$) for the subset of the final sample with \hst~counterpart detections and that is complete down to \Mstar\,$=$\,$10^7$\,\Msun. The data points are color-coded by stellar core mass. When omitting the most massive stellar core at a large offset, a positive correlation is observed. This result may indicate that only the most massive stellar cores experience significant orbital evolution.}}
\label{fig:PETRORADR_DELTAS_MSTAR}
\end{figure}

\section{HLX Occupation Fraction}
\label{sec:Occ}

Numerical simulations predict that Milky Way-mass halos contain an average of $\sim$10 wandering massive black holes due to mergers, with implications for the demographics of massive black holes and potential binary black hole formation \citep{Tremmel:2018b,Ricarte:2021b}. HLXs are unique tracers of this population, and Figure \ref{fig:Occ_Frac_MSTAR_LX} shows the number of HLXs per galaxy as a function of X-ray luminosity from our sample. These occupation fractions are measured by, in each luminosity bin, finding all parent galaxies in the CSC footprint with the sensitivity to measure down to the lower luminosity limit of the bin. In each bin, the occupation fraction is the number of HLXs (corrected for the estimated contamination fraction; Section \ref{sec:XRay}) over the total number of parent sample galaxies. We have further divided these fractions into subsets based on their \MstarTwo~values (accounting for the mass sensitivity limits). The occupation fraction declines nearly an order of magnitude over the full mass range of our sample, indicating that the vast majority of HLXs found will be in relatively low-mass stellar cores. 

For our full sample of HLXs, the observed occupation fraction is $\sim$\,$10^{-2}$. To correct for the fraction of accreting black holes with X-ray luminosities below the $10^{41}$\,\uLum~luminosity threshold, we divide the overall fraction of HLXs in our sample by the cumulative fraction of X-ray AGN at \LXB\,$=$\,$10^{41}$\,\uLum~(from \citealt{Lehmer:2008}; based on AGN detected down to 2$-$8\,keV luminosities of $\sim$ $10^{39}$\,\uLum~and converted to rest-frame 0.5$-$7\,keV assuming a powerlaw spectrum defined by $\Gamma$\,$=$\,1.7). This yields a corrected occupation fraction of $\sim$\,$10^{-1}$. If we further incorporate the host galaxy stellar mass dependence of the X-ray AGN fraction from \citet{Lehmer:2008} (which accounts for the declining fraction of AGN with decreasing host galaxy mass over the range \Mstar\,$=$\,$10^{9.5-11}$\,\Msun), the corrected occupation fraction becomes $\sim$10 per galaxy and is in-line with theoretical predictions \citep{Tremmel:2018b,Ricarte:2021b,Ricarte:2021a}. However, the possibility of tidal stripping means the original stellar core masses may have been larger, in which case the true occupation fraction will be smaller. Potential implications of lower occupation fractions would be fewer seed massive black holes and/or less frequent accretion of them onto halos than predicted.

Furthermore, using the host galaxy SED classifications for our sample (Table \ref{tab:Samp}), we find that the relative fractions of early-type host galaxies (based on red and passive spectroscopic signatures) and late-type host galaxies (based on blue and star forming spectroscopic signatures) in our sample are consistent, to within less than 1$\sigma$, with the relative fractions in the parent sample from which they were drawn (same sample used in Figure \ref{fig:Occ_Frac_MSTAR_LX}). This result is actually consistent with numerical work predicting that the number of wandering massive black holes in galaxy halos is not strongly dependent on galaxy morphology due to the large timescales over which satellites are accreted \citep{Tremmel:2018b}.

The stellar cores hosting HLXs may evolve toward smaller nuclear offsets via dynamical friction, which is dependent on the stellar core mass \citep[e.g.,][]{Boylan-Kolchin:2008}. This effect is observationally confirmed in our sample from a negative correlation between the HLX stellar core masses and projected physical offsets (Spearman rank correlation probability of \MstarDeltaSSpearP~among the subset showing morphological evidence for interaction with the host galaxy; Figure \ref{fig:DELTAS_MSTAR_BIN}). While this correlation may be affected by a possible bias against finding low-mass stellar cores at small separations, the most massive stellar cores still exhibit a strong preference for small physical offsets. However, Figure \ref{fig:PETRORADR_DELTAS_MSTAR} also shows how the projected physical offsets may have a slight positive correlation (Spearman rank probability of \DeltaSPetroRadSpearStat) with the physical characteristic size of the host galaxies (parameterized by the $r$-band Petrosian radii). This may reflect that, aside from the HLXs in the most massive stellar cores, many of the HLX offsets may be weakly affected by dynamical friction and do not experience significant evolution toward smaller separations.

Continued orbital migration will lead to the formation of bound binaries with the host galaxy central massive black holes, and hence future gravitational wave sources detectable by pulsar timing arrays \citep[e.g.,][]{Hobbs:2010,Arzoumanian:2020} and by the planned Laser Interferometer Space Antenna mission \citep[LISA; e.g.,][]{Amaro-Seoane:2017}. The probability of forming a massive black hole binary is predicted to increase with host galaxy stellar mass and (more strongly) with the merger mass ratio \citep{Tremmel:2018b}. Applying these predictions to our sample suggests binary formation for $\sim$\,5\% of the HLXs. A corollary of this result is that the less massive stellar cores are more likely to remain wandering at larger separations and will constitute the majority of the black hole mass distribution within galaxy halos.

\section{Conclusions}
\label{sec:conc}

We systematically identify 24 HLX candidates that have archival \hst~imaging. We use the high-resolution and sensitive imaging to understand the nature of the accreting X-ray sources, identify potential IMBH candidates, and offer constraints on the HLX occupation fraction in galaxies and implications for wandering massive black holes in galaxy halos. Our primary conclusions are as follows: \\ \\

\begin{enumerate}

\item Of the \Samp~HLX candidates, \hst~counterparts are detected for 23 of them. The \hst~imaging reveals morphological evidence for interaction between the HLX candidates and their host galaxies in 12 cases. After removing known contaminants, $\sim$70\% of the HLX candidates are expected to be true HLXs. \\ \\

\item Based on the host galaxy redshifts, stellar masses for at least 21 of the HLX candidate \hst~counterpart detections are $\geq$\,$10^7$\,\Msun, exceeding that of the most massive globular clusters observed. This suggests that at least 88\% of true HLXs are in stellar cores or stellar core remnants from galaxies that merged with the host galaxy. \\ \\

\item The remaining 12\% of the sample could be IMBHs that formed in-situ within globular clusters, or exceptionally luminous XRBs. For the HLX candidate without an \hst~counterpart detection, if it resides in a remnant stellar core then significant tidal stripping likely occurred. \\ \\

\item The HLX candidates show evidence for relatively high X-ray luminosities per stellar core mass compared to X-ray AGN located in galaxy nuclei (median offset of \MedianLXMstarOffsetDex~dex), primarily driven by the lower mass stellar cores. If this offset represents enhanced HLX luminosities, it may suggest that galaxy mergers are an efficient route for growing IMBHs. \\ \\

\item The stellar cores with the smallest projected physical offsets from the host galaxy nuclei are biased toward relatively large masses (Spearman rank correlation probability of \MstarDeltaSSpearP), and the physical offsets are mildly correlated with the host galaxy physical extents. These results are consistent with the expectation that relatively massive stellar cores more efficiently evolve toward smaller separations through dynamical friction. By implication, the majority of wandering massive black holes in galaxy halos will be in low-mass stellar cores and hence are likely IMBHs.

\item The average Eddington ratio of the HLXs is \FEddFitScalRelaAGN~assuming their stellar counterparts follow a scaling relation between nuclear massive black holes and host galaxy stellar masses for AGN. After correcting for contamination, this assumption further predicts that, if the black hole mass estimates are at their lower limits, then up to \IMBHPercCorrAGN\% of the HLXs in our sample are IMBHs. The X-ray-to-optical emission ratios are also consistent with a significant fraction of the sample being associated with black holes less massive than those of typical AGN.

\item The estimated HLX occupation fraction in galaxies is $\sim$\,$10^{-2}$. After correcting for the HLX selection luminosity threshold and AGN dependence on host galaxy mass, the HLX galaxy occupation fraction is $\sim$\,0.1\,$-$\,10. These estimates are potentially consistent with numerical predictions of $\sim$\,10 wandering massive black holes per Milky Way-mass halo. Extrapolation of numerical predictions suggests $\sim$\,5\%
of the HLX sample will form massive black hole binaries.

\end{enumerate}

\begin{acknowledgments}

We thank an anonymous reviewer for the detailed and insightful comments that have greatly improved the manuscript quality. The results reported here are based on observations made with the NASA/ESA \emph{Hubble Space Telescope}, obtained at the Space Telescope Science Institute, which is operated by the Association of Universities for Research in Astronomy, Inc., under NASA contract NAS5-26555. This research has made use of data obtained from the \ch~Data Archive and the \ch~Source Catalog, and software provided by the \ch~X-ray Center (CXC) in the application packages \ciao~and \sherpa. This work was supported by the \hst~archival project HST-AR-15787.001-A. MM acknowledges supported from the program Unidad de Excelencia María de Maeztu CEX2020-001058-M. M.M. acknowledges support from the Spanish Ministry of Science and Innovation through the project PID2021- 124243NB-C22. The work of DS was carried out at the Jet Propulsion Laboratory, California Institute of Technology, under a contract with NASA. Some of the data presented in this article were obtained from the Mikulski Archive for Space Telescopes (MAST) at the Space Telescope Science Institute. The specific observations analyzed can be accessed via \dataset[DOI: 10.17909/bv1y-zp75]{https://doi.org/10.17909/bv1y-zp75}.

\end{acknowledgments}

\facilities{\hst, Sloan, CXO}

\software{\astropy\footnote{\href{\astropylink}{\astropylink}} \citep{astropy:2013, astropy:2018,astropy:2022}, \texttt{CIAO} \citep{Fruscione:2006}, \texttt{Sherpa} \citep{Freeman:2001,Doe:2007,Burke:2020}}

\end{document}